# A Dual-Head Transformer–State-Space Architecture for Neurocircuit Mechanism Decomposition from fMRI

Cole Korponay

January 16, 2026

## Abstract

Precision psychiatry aspires to elucidate brain-based biomarkers of psychopathology to bolster disease risk assessment and treatment development. To this end, functional magnetic resonance imaging (fMRI) has helped triangulate brain circuits whose functional features are correlated with or even predictive of forms of psychopathology. Yet, fMRI biomarkers to date remain largely descriptive identifiers of where, rather than how, neurobiology is aberrant, limiting their utility for guiding treatment. We present a method for decomposing fMRI-based functional connectivity (FC) into constituent biomechanisms – output drive, input responsivity, modulator gating – with clearer alignment to differentiable therapeutic interventions. Neurocircuit mechanism decomposition (NMD) integrates (i) a graph-constrained, lag-aware transformer to estimate directed, pathway-specific routing distributions and drive signals, with (ii) a measurement-aware state-space model (SSM) that models hemodynamic convolution and recovers intrinsic latent dynamics. This dual-head architecture yields interpretable circuit parameters that may provide a more direct bridge from fMRI to treatment strategy selection. We instantiate the model in an anatomically and electrophysiologically well-defined circuit: the cortico-basal ganglia-thalamo-cortical loop.

## Introduction

Functional connectivity (FC) has been valuable for identifying which brain circuits appear aberrant in a given disorder, but it provides limited insight into how or why circuit interactions differ in mechanistic terms. A single FC value can reflect multiple underlying processes—changes in upstream drive, changes in downstream sensitivity, or changes in gating by an external modulator. These possibilities have different clinical implications. (Biswal et al., 1995; Power et al., 2011; Yeo et al., 2011)

Here, we formalize an integrated, mechanism-first architecture intended to deliver a single consistent decomposition of circuit behavior into drive, timing, baseline responsiveness, dynamic gating, and stability. The key innovation is reframing functional connectivity as a multivariate routing problem. Rather than summarizing circuit coupling with pairwise correlations, the model assigns each target region's activity to a competitive mixture of candidate sources, estimating which regions drive which targets and at what delays. Because contributions are learned in the context of all other regions' competing explanatory power, the method distinguishes pathway-specific influence from correlations that arise indirectly via shared inputs or global fluctuations. This yields interpretable, anatomically constrained features—directed pathway strength, delay profiles, and state-dependent modulation—that are more directly mappable to circuit hypotheses than conventional FC. (Alexander et al., 1986; Haber, 2016; Friston et al., 2003)

Below is a summary of the pipeline at a high level:

- Stage 0 — ROI specification & circuit graph: define circuit nodes (ROIs/parcels) and a directed adjacency mask of allowed anatomical edges. (Schaefer et al., 2018; Yeo et al., 2011; Alexander et al., 1986)
- Stage 1 — Data conversion: extract ROI time series (region × time), optionally sub-parcellate key structures (e.g., striatum K-parcellation), and save standardized inputs (e.g., NPZ).
- Stage 2 — SSM deconvolution: infer deconvolved latent neural state estimates x-hat(t) while explicitly modeling hemodynamics (HRF) and measurement noise. (Glover, 1999; Handwerker et al., 2004; Woolrich et al., 2004)
- Stage 3 — Transformer routing: operate on x-hat(t) to infer directed, lag-resolved routing tensors and pathway-level drive signals under anatomical constraints. (Vaswani et al., 2017; Shaw et al., 2018)
- Stage 4 — Feature extraction: compute pathway influence, delay statistics (peak/centroid/dispersion), baseline sensitivities, and intrinsic dynamics descriptors. (Kalman, 1960; Friston et al., 2003; Gu et al., 2015)
- Stage 5 (optional) — Modulatory gating: estimate time-varying gains g(t) applied within the SSM to capture state-dependent routing modulation.
- Stage 6 — Downstream analysis: use mechanistic features for subject-level inference, clustering/subtyping, and phenotype association.

# Methods

## 5. Aim 1 Methods: Transformer–SSM Model Architecture

### 5.1 Shared temporal encoder

Both model heads operate on a common set of token embeddings derived from region-specific BOLD time series. A lightweight temporal encoder—implemented as compact 1D convolutions or learned temporal filters—extracts local temporal features from windowed segments of each region's signal. Tokens preserve temporal ordering and regional identity. This shared representation reduces redundant computation while allowing each head to specialize: the transformer focuses on cross-region predictive relationships (propagation), while the SSM focuses on within-state dynamics and measurement modeling. (Vaswani et al., 2017)

Formally, for each region j at time t, we construct an embedding $e_j(t) = Enc\_psi(y_j(t-\Delta:t))$, where Enc_psi is a small temporal feature extractor applied to a fixed-length history window of that region's signal. (Kim et al., 2021)

## 5.2 Head 1: Graph-constrained transformer for signal propagation modeling

The transformer head models signal propagation through the circuit by learning directed, lag-aware predictive dependencies between nodes while respecting known anatomical constraints. A hard anatomy mask restricts attention to biologically plausible source→target edges (e.g., cortex→striatum permitted; direct striatum→cortex projections disallowed), shrinking the parameter space and improving interpretability. (Vaswani et al., 2017; Kan et al., 2022)

### *Factorized spatial–temporal attention*

To reduce computational complexity and enhance interpretability, attention is factorized into separable spatial and temporal components. Spatial attention learns which source regions predict each target region's activity within a lag bucket (directed, graph-masked). Temporal attention learns which relative delays are most predictive (0–12 s in TR-width buckets, with learned relative lag embeddings). For a target region j, spatial logits $S_{ij}$ are formed over sources i and temporal logits $T_l$ are formed over lags l. These are combined additively in logit space as $logit_{\{i,l \rightarrow j\}} = S_{ij} + T_l$, and a single softmax is applied over all (source, lag) keys to produce a probability-like routing tensor $\pi(i,l \rightarrow j)$. (Shaw et al., 2018)

### *Transformer outputs and derived metrics*

- Edge-specific influence weights: strength of predictive contribution for each permitted pathway i→j.
- Lag statistics per pathway: peak lag (most predictive delay), centroid (weighted average delay), and concentration (inverse variance; timing precision).
- Per-target aggregated drive signals u-hat(t): attention-weighted sums of lagged source activity that serve as pathway-level inputs.

The pathway-level drive for target j at time t is computed as $\hat{u}^j(t) = \sum_{i,l} \pi(i,l \rightarrow j) \cdot x_i(t-l)$, where $x_i$ denotes deconvolved neural-state estimates provided by the SSM.

## 5.3 Head 2: Measurement-aware state-space model for dynamics

The SSM head recovers latent neural dynamics from observed BOLD signals while explicitly modeling the hemodynamic response as a known observation process. This separation is essential because apparent temporal patterns in BOLD may reflect vascular delays rather than neural propagation. (Handwerker et al., 2004; Glover, 1999; Woolrich et al., 2004)

### *Dynamics model*

Latent neural state evolution follows a discrete-time linear state-space formulation: $x(t+1) = A x(t) + B \hat{u}(t) + w(t)$. Here x(t) represents latent neural activity states (one per ROI), A captures intrinsic dynamics (state persistence and recurrent coupling), B maps pathway-level drive inputs to state changes, and w(t) represents process noise. (Kalman, 1960)

### *Interpretation of baseline sensitivity B and clinical caution*

The B matrix quantifies pathway-specific input sensitivity during observed spontaneous dynamics—how strongly each pathway responds to its endogenous drive. Anatomical constraints ensure B has non-zero entries only for plausible input pathways. Importantly, B reflects sensitivity observed during natural activity and may differ from responsiveness to external interventions (e.g., TMS, DBS, pharmacotherapy). Accordingly, dynamics metrics are treated as hypothesis-generating mechanistic features, not validated predictors of treatment response.

### Biological plausibility constraints
- Spectral stability: eigenvalues of A constrained to lie within the unit circle to prevent explosive dynamics.
- Pathway-aligned sparsity: entries in A and B constrained by anatomical masks, reducing degrees of freedom.
- Low-rank input structure: B parameterized via low-rank factorization (e.g., rank-3) to reduce overfitting and encourage interpretable input modes.

### Observation model with HRF convolution
The relationship between latent neural states and observed BOLD signals incorporates hemodynamic convolution: $y(t) = (h * Cx)(t) + v(t)$, where h denotes a region-specific hemodynamic response function (HRF) and $v(t)$ captures measurement noise (e.g., AR(1) plus white noise). HRF parameters are constrained by physiological priors (positive peak, plausible time-to-peak and dispersion, bounded undershoot). (Woolrich et al., 2004; Handwerker et al., 2004)

### Inference and outputs
Inference is performed using differentiable Kalman filtering/smoothing (Backprop-KF) to enable end-to-end optimization. The SSM produces deconvolved state estimates $\hat{x}(t)$, subject/region HRF parameters, and physical-unit dynamics descriptors including characteristic time constants, damping ratios, stability metrics, and controllability indices derived from the (A,B) pair along specified routes. (Haarnoja et al., 2016)

## 5.4 Coupling strategies: deconfounding and integration (Friston et al., 2003)
A key architectural decision is how the transformer and SSM interact. The goal is separate identifiability of transformer-derived drive, transformer-derived timing, SSM-derived baseline sensitivity, and SSM-derived modulatory gating, while preventing circular parameter trade-offs. To achieve this, we use explicit deconfounding via asymmetric stop-gradient coupling. The SSM first deconvolves BOLD into $\hat{x}(t)$. The transformer operates on $\hat{x}(t)$ (not raw BOLD) to infer directed, lag-resolved routing and compute drive $\hat{u}(t)$. Gradients do not pass back through the SSM (stop-gradient at the junction).

### Midbrain-derived modulatory gating
To capture state-dependent routing, a modulatory gate $g(t)$ is computed from a designated modulatory subsystem (e.g., SN/VTA midbrain signals) and applied inside the SSM. This produces an effective sensitivity $B\_eff(t) = B(I + \beta * G(t))$, preserving interpretability of baseline sensitivity B versus modulatory gating $g(t)$.

## 5.5 Hyperparameters and optimization
To prevent overfitting through architecture search, core architectural parameters are pre-registered based on interpretability goals, fMRI constraints, and best practices.

*Transformer parameters*
- Depth: one attention block (no stacking) to keep attribution (who/when) directly tied to inferred routing.
- Heads: 4 total, partitioned as 2 spatial heads and 2 temporal heads.
- Hidden dimension: 128; per-head Q/K/V dimension: 32.
- Temporal attention span: 0–12 s in TR-width buckets with learned relative lag embeddings. (Shaw et al., 2018)
- Fusion: additive logits with a single softmax over (source × lag).
- Hard anatomical graph mask to eliminate implausible edges.
- Attention-only transformer (no MLP block), Pre-LayerNorm, dropout = 0.1 on attention probabilities. (Xiong et al., 2020)

*SSM parameters*
- Dynamics/observation: $x_{t+1} = A x_t + B u_t + w_t$; $y_t = HRF * (C x_t) + v_t$. (Kalman, 1960)
- A: sparse and graph-masked with stability constraint $\rho(A) \leq 0.98$; fixed L1 penalty on allowed entries.
- B: time-invariant, low-rank factorization (e.g., rank-3); C set to identity for transparency.
- HRF: region-specific canonical basis with constrained variability; support 0–20 s.
- Noise covariances: diagonal Q and R shared across subjects; initial state $x_0 = 0$.
- Optional multiplicative gating computed from modulatory subsystem and bounded in amplitude.

*Optimized hyperparameters*
- Learning rate (Adam/AdamW): grid {3e-4, 1e-4, 3e-5}. (Kingma & Ba, 2015)
- Coupling choice Strategy 1 vs Strategy 2: Strategy 2 adopted only if validation NLL improves by at least 0.3% and identifiability/reliability gates pass.

## 5.6 Optimization strategy: losses, priors, and identifiability

Training optimizes a composite objective balancing reconstruction accuracy, biological plausibility, and identifiability.

**Primary loss:** Negative log-likelihood of observed BOLD under the SSM, computed via Kalman filtering/smoothing.

**Transformer auxiliary losses:** Multi-step forecasting (2–4 TR ahead) to emphasize predictive dependencies rather than instantaneous correlations, plus optional reconstruction losses at the last frame.

**Regularization and priors:** L1 sparsity on routing weights; smooth/unimodal lag priors; physiological priors on HRF timing and shape; stability constraints on A.

**Timescale-based identifiability:** Neural routing and propagation are encouraged to occupy fast lag ranges (short delays) while vascular convolution is modeled separately as a smoother, slower observation process. (Mitra et al., 2014; Handwerker et al., 2004)

# Summary of key output metrics (FC decomposition)

Table 1 summarizes the key output metrics produced by the framework and how they relate to traditional static functional connectivity (FC) in the context of corticostriatal circuitry. In particular, the method decomposes apparent corticostriatal FC into separable components corresponding to (i) cortical drive patterns, (ii) striatal receptivity (baseline input sensitivity / explainable variance), and (iii) midbrain gating.

| Output metric | How it is computed (conceptually) | What it represents | Why it matters clinically |
|---|---|---|---|
| Traditional FC (baseline reference) | Pairwise correlation or covariance between ROI time series (e.g., Corr($C_i(t)$, $S(t)$)). | Undirected statistical coupling that conflates drive, timing, and shared confounds. | Common biomarker input, but hard to map onto pathway-specific intervention targets. |
| Cortical drive profile (static) | Multivariate mapping from cortex to striatum, e.g., $S(t) \approx \Sigma_i w_i C_i(t)$. Drive weights are $\{w_i\}$. | Which cortical sources contribute most to striatal fluctuations (source composition). | Identifies candidate upstream sources to target (e.g., stimulation or cognitive control nodes). |
| Striatal receptivity (static) | Scalar sensitivity to cortical input, e.g., $R^2$ of $S(t)$ predicted by cortex, or ‖Cov(C,S)‖ magnitude. | How strongly striatum expresses cortical influence overall (net coupling amplitude). | Potential stratifier: high vs low receptivity may indicate different circuit regimes or treatment responsiveness. |
| Directed pathway influence (edge strength) | From routing tensor $\pi(i,\ell \to j)$, influence strength for $i \to j$ is $\Sigma_\ell \pi(i,\ell \to j)$. | Estimated directed contribution of a specific anatomical pathway. | Enables pathway-level hypotheses and comparative analysis across circuit branches. |
| Lag timing metrics (peak / centroid / concentration) | From $\pi(i,\ell \to j)$ over lag buckets $\ell$: peak lag = argmax; centroid = weighted mean; concentration = inverse variance. | When influences arrive (delay profile) and how temporally precise they are. | Timing signatures can distinguish fast vs slow pathways and separate neural routing from vascular delay. |
| Drive signal $\hat{u}(t)$ (time-resolved input) | Attention-weighted sum of lagged sources into each target: $\hat{u}^j(t)=\Sigma_{i,\ell} \pi(i,\ell \to j) x_i(t-\ell)$. | Moment-to-moment endogenous input delivered to each target node. | Provides time series features for state detection and relating circuit input to symptoms/behavior. |
| Baseline input sensitivity B (SSM parameter) | In SSM: $x(t+1)=A x(t)+B \hat{u}(t)+w(t)$. B captures time-invariant sensitivity to drive inputs. | How responsive each target is to its endogenous inputs during spontaneous dynamics. | Mechanism-like parameter that can support patient stratification and intervention hypothesis generation. |
| Dynamic modulation / gating $g(t)$ | Time-varying gain applied to selected pathway families inside SSM: $\hat{u}_{eff}(t)=(1+\beta g(t)) \hat{u}(t)$. | State-dependent amplification or suppression of routing (contextual gating). | Captures circuit state changes that may reflect arousal, control, neuromodulation, or symptom-relevant switching. |

# Code Availability

https://github.com/ckorponay/neurocircuit-mechanism-decomposition